# Emergent chiral symmetry breaking in moiré domain wall networks redirects topological boundary states in bilayer graphene

Xue Yan[1], Kaiyun Chen[2], Yuan Yan[3], Fan Feng[1], Minglei Sun[1], Christian Brandl[1], Jefferson Zhe Liu[1*]

[1]Department of Mechanical Engineering, The University of Melbourne, Parkville, VIC 3010, Australia
[2]Northwest Institute for Nonferrous Metal Research, Xi'an 710016, China
[3]School of Chemical Engineering, University of New South Wales, Sydney, NSW 2052, Australia

Lattice-mismatched bilayer graphene self-organizes into a moiré network of one-dimensional domain walls that conduct electrons with low dissipation, attractive for low-power electronics. We show these channels are not always straight: as the lattice relaxes to minimize strain, the network can spontaneously curve into chiral morphology. Atomistic simulations map a phase diagram in which the strain–flexibility balance selects one of three stable domain wall morphologies — straight, mono-chiral, or dual-chiral. Electronic structure calculations show that this morphology controls where low-energy electrons accumulate: straight channels concentrate states at the domain-wall-connecting nodes, while chiral channels shift that weight onto the domain walls themselves. This network geometric switch lets the same moiré material support either localized electronic hot spots or directional conducting channels — two strategies for guiding electrons in low-power graphene devices.

*Introduction*—In graphene bilayers, twist or strain mismatch generates long-wavelength moiré superlattices [1,2] that relax into distinct stacking-symmetry domains separated by topological domain walls (TDWs) [3,4]. These TDWs typically interconnect to form a complex moiré TDW network connected through a set of nodes, known as topological crossing points (TCPs). Opposite valley Chern numbers across neighboring domains endow these walls with one-dimensional topologically protected electronic boundary states [5–7], making TDWs central to the low-energy electronic structure of relaxed moiré graphene [4,6,8–11]. While the stacking-domain topology guarantees the existence of these channels [12–15], it does not determine the TDW geometry in the network [16,17]. Experiments have revealed either straight or strongly curved, swirl-like TDWs across diverse moiré systems [3,18–24], indicating that such symmetry breaking in network geometry constitutes an additional degree of freedom [24,25]. Exploiting this extra degree of freedom for low-energy electronics requires understanding how this network-geometry symmetry breaking emerges under a fixed stacking-domain topology and how it reorganizes the electronic boundary states.

The TDWs are not only topological interfaces but also elastic line defects. They can be rigorously described as partial basal-plane dislocations characterized by a Burgers vector associated with interlayer translation [26-28]. Classical dislocation theory shows that the energy of a line defect depends on its orientation relative to the Burgers vector [28-30], implying that the strain mismatch, naturally existing in the moiré system, plays a critical role. An isolated TDW could tend to curve, increasing its screw-like character and lowering its elastic energy. Note that in a connected moiré network, TDWs are constrained by TCP connectivity and the spacing imposed by the superlattice periodicity [29]. The resulting network geometry should emerge from a competition between orientation-dependent wall/dislocation energetics and the connectivity constraints required by the superlattice periodicity.

In this letter, we show that lattice relaxation drives such network geometric symmetry breaking. Atomistic calculations reveal a strain–flexibility phase diagram comprising straight, mono-chiral, and dual-chiral equilibrium networks. Large-scale tight-binding calculations further show that chirality reorganizes their low-energy electronic states: straight networks concentrate spectral weight near TCPs, whereas chiral networks redistribute it toward TDWs and produce directional edge asymmetry. Network chirality therefore provides a geometric degree of freedom that selects between TCP-localised and wall-directed low-energy states in moiré graphene.

*Results*—To determine how mechanical relaxation shapes TDW network geometry, we relaxed strained graphene moiré bilayers across a two-dimensional parameter space defined by biaxial misfit strain and bottom-layer flexibility. The bottom flexibility is controlled by either allowing or constraining out-of-plane relaxation of the bottom layer, thereby corresponding to the substrate-pinned (*z*-rigid) or free-standing bilayers (*z*-free) as in most experiments (Supplemental Material [36] and Fig. S1-S2). Figure 1(a) summarises the resulting equilibrium structures, visualized using our previously developed stacking-order index, which maps the local stacking order across the network [30]. In the unrelaxed moiré case, AB and BA stacking domains are separated by TDWs that intersect at AA-stacking TCPs [29]. After relaxation, the TDW network is no longer confined to this idealized straight grid: the walls bend collectively and develop curved morphology. Around each TCP, the bending of neighboring walls defines a local handedness, which provides a natural basis for classifying network chirality.

*Contact author: zhe.liu@unimelb.edu.au

Three distinct equilibrium morphologies emerge within this strain–flexibility space: a straight network that preserves the underlying lattice symmetry and two chiral networks, termed mono-chiral and dual-chiral, that break mirror and inversion symmetry (Fig. 1(a)). In mono-chiral networks, all TCPs share the same handedness, whereas in dual-chiral networks, left- and right-handed TCPs alternate across the lattice. The phase sequence depends strongly on the bottom layer's flexibility. In the *z*-rigid case, increasing strain stabilizes mono-chiral networks at 0.1%, followed by dual-chiral networks at 0.2–0.24%. In the *z*-free case, the sequence reverses: dual-chiral networks appear at low strain, mono-chiral networks at intermediate strain, and nearly straight walls re-emerge at higher strain. This contrast is evident at fixed strain: for example, at 0.24%, the z-rigid system remains dual-chiral, whereas the z-free system shows strongly suppressed chirality. These results demonstrate that TDW network geometry is not uniquely determined by stacking topology but instead arises from the interplay between in-plane strain and the bottom layer's flexibility.

To quantify the TDW geometry, we introduce two descriptors. The chiral angle α measures the deviation of a TDW from the line connecting neighboring TCPs (Fig. 1(b)-(c)), and the curvature localization parameter $R_c$ characterizes the spatial concentration of bending along a wall (Fig. 1(d)). In mono-chiral networks, α is finite and nearly uniform for all TDWs within a given network, indicating consistent handedness across the lattice. In dual-chiral networks, curvature localizes between neighboring TCPs with identical handedness, while walls connecting the neighboring TCPs with opposite handedness remain nearly straight. In straight networks, both α and $R_c$ approach zero. To distinguish the two different types of TDWs in the mono-chiral and due-chiral cases, we also introduce two additional chiral angle descriptors α' and α'' as shown in Fig. 1(b)-(c).

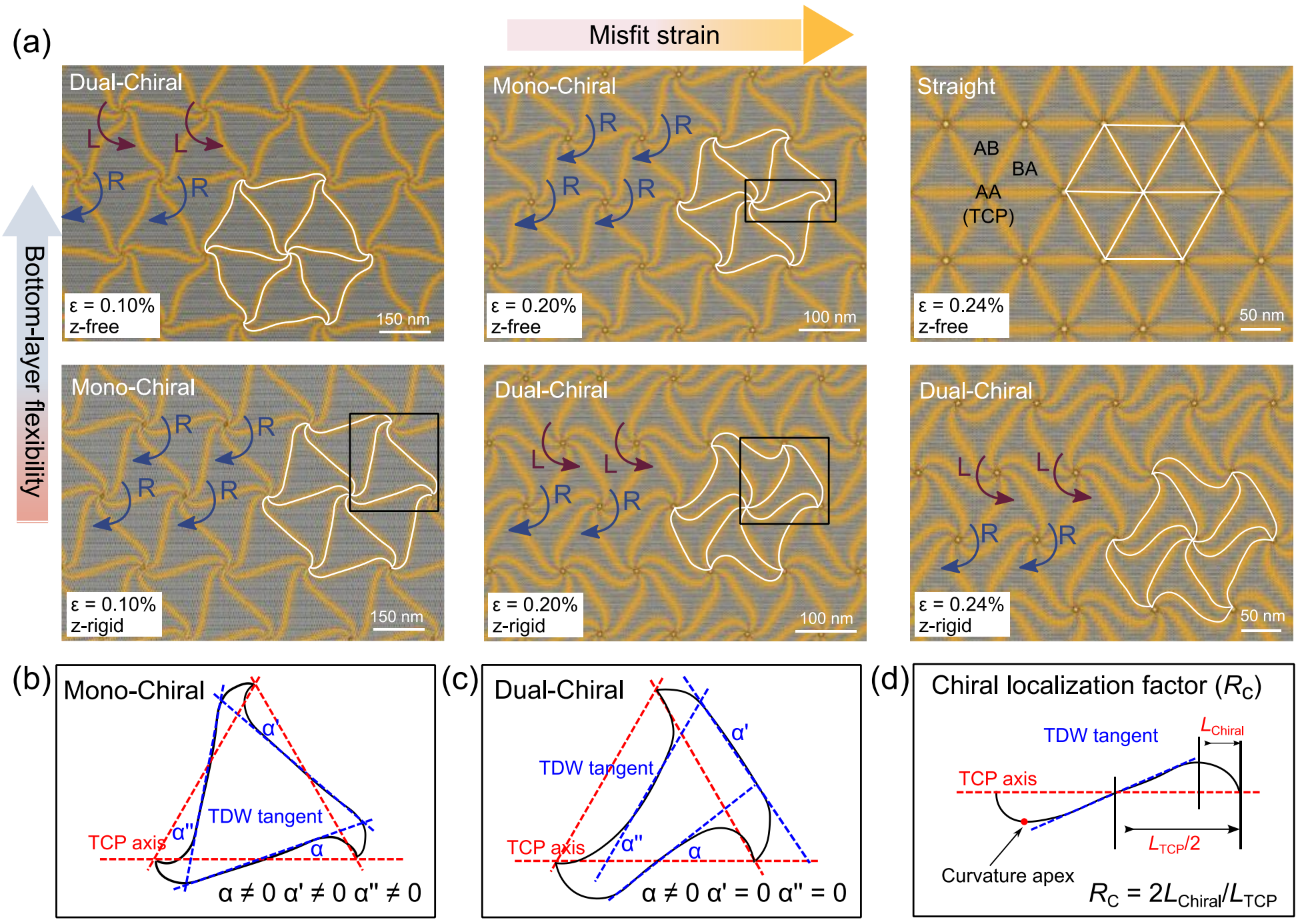

FIG 1. Classification of moiré TDW network geometries and geometric descriptors of chirality. (a) Representative TDW morphology with the variation of biaxial misfit strain and bottom-layer flexibility. Stacking-order index [30] identifies AB/BA domains (grey), TDWs (yellow), and AA-stacked TCPs. (b,c) Chiral angle α, defined by the TDW deviation from the TCP–TCP axis at its midpoint. (d) Curvature localization $R_c$, the apex-to-nearest-TCP distance normalized by half the TCP–TCP separation.

Using these descriptors, we construct the phase diagram in Fig. 2(a) as a function of biaxial misfit strain and bottom-layer flexibility, quantified by the inverse stiffness $1/k$ [36]. Marker color represents the chiral angle α, and the angular sector represents the curvature localization parameter $R_c$. For each parameter set, the lowest-energy morphology is selected [36]. Two main trends emerge from this phase map. At large misfit strain ($\varepsilon > 0.83\%$), TDWs adopt the straight morphology across the full range of flexibility, whereas at small misfit strain ($\varepsilon < 0.23\%$), only chiral morphologies are stable, where mono-chiral and dual-chiral networks appear in similar regions of the phase diagram (Fig. S3). Between these two regions, bottom-layer flexibility plays a decisive role in determining whether chirality persists. Highly flexible systems (bottom layer) transition to straight TDWs at $\varepsilon \approx 0.23\%$, whereas rigid systems require substantially larger misfit strain ($\varepsilon \approx 0.83\%$) to suppress chirality. At a fixed misfit strain, increasing flexibility can, therefore, change not only the chirality boundary, but also the sequence of geometric transitions, including dual-chiral → mono-chiral → straight (e.g., $\varepsilon = 0.30\%$) or direct dual-chiral → straight transitions (e.g., $\varepsilon = 0.23\%$) (Fig. S4).

*Contact author: zhe.liu@unimelb.edu.au

Within the chiral regime, the geometric descriptors $\alpha$ and $R_c$ change systematically across the phase diagram. Both quantities initially increase with misfit strain and subsequently decrease as the morphology approaches a straight one. For example, along the z-rigid line ($k = 0^+$), $R_c$ increases from $\varepsilon \approx 0.1\%$ to ~0.8%, accompanied by a deepening of the dual-chiral (red) or mono-chiral (blue) color that reflects increasing α. Beyond $\varepsilon \gtrsim 0.8\%$, both $R_c$ and α collapse to zero as the system enters the straight-TDW regime. This behavior reflects a progressive localization and eventual suppression of curvature. Upon crossing into the straight-TDW regime, $\alpha$ and $R_c$ drop abruptly to zero, indicating an abrupt loss of geometric chirality. Note that the variation of $\alpha$ and $R_c$ with respect to misfit strain and flexibility is not monotonic, particularly for $\varepsilon <$ ~0.2%. Overall, increasing misfit strain suppresses TDW chirality, while greater bottom-layer flexibility lowers the strain needed to form straight walls.

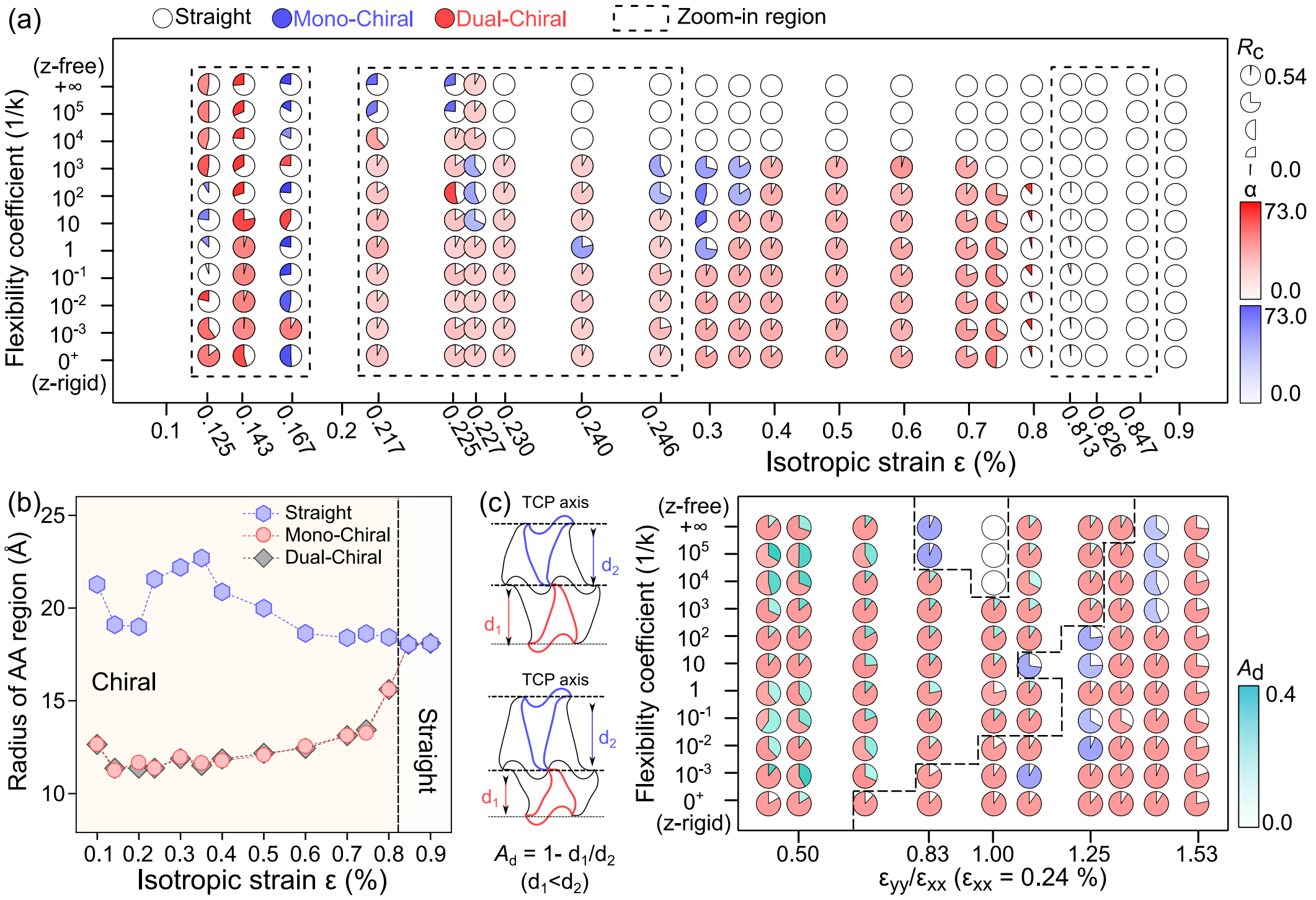

FIG 2. Phase diagram of moiré TDW network geometry. (a) Ground-state geometries in biaxial misfit strain–flexibility space, with flexibility defined as $1/k$, where $k$ is the stiffness. (b) AA-region (TCP) size for the lowest-energy geometry, showing sensitive morphology dependence at fixed misfit strain. (c) Left: definition of AB/BA domain-size symmetry and asymmetry descriptor $A_d$. Right: phase map versus anisotropic misfit strain and flexibility.

Figure 2(b) shows that chirality is accompanied by a pronounced reduction in the size of the AA-stacked region (*i.e.*, TCPs). At the same misfit strain, the straight-wall morphology exhibits the largest AA domain size, with a radius of about 20 Å, whereas both mono-chiral and dual-chiral networks significantly reduce the AA domain size (down to ~10 Å). The similar AA-domain sizes in the two chiral phases indicate that this reduction is driven mainly by curvature rather than by handedness. As misfit strain increases and chirality weakens (Fig. 2(a)), the contrast in AA size correspondingly diminishes.

*Contact author: zhe.liu@unimelb.edu.au

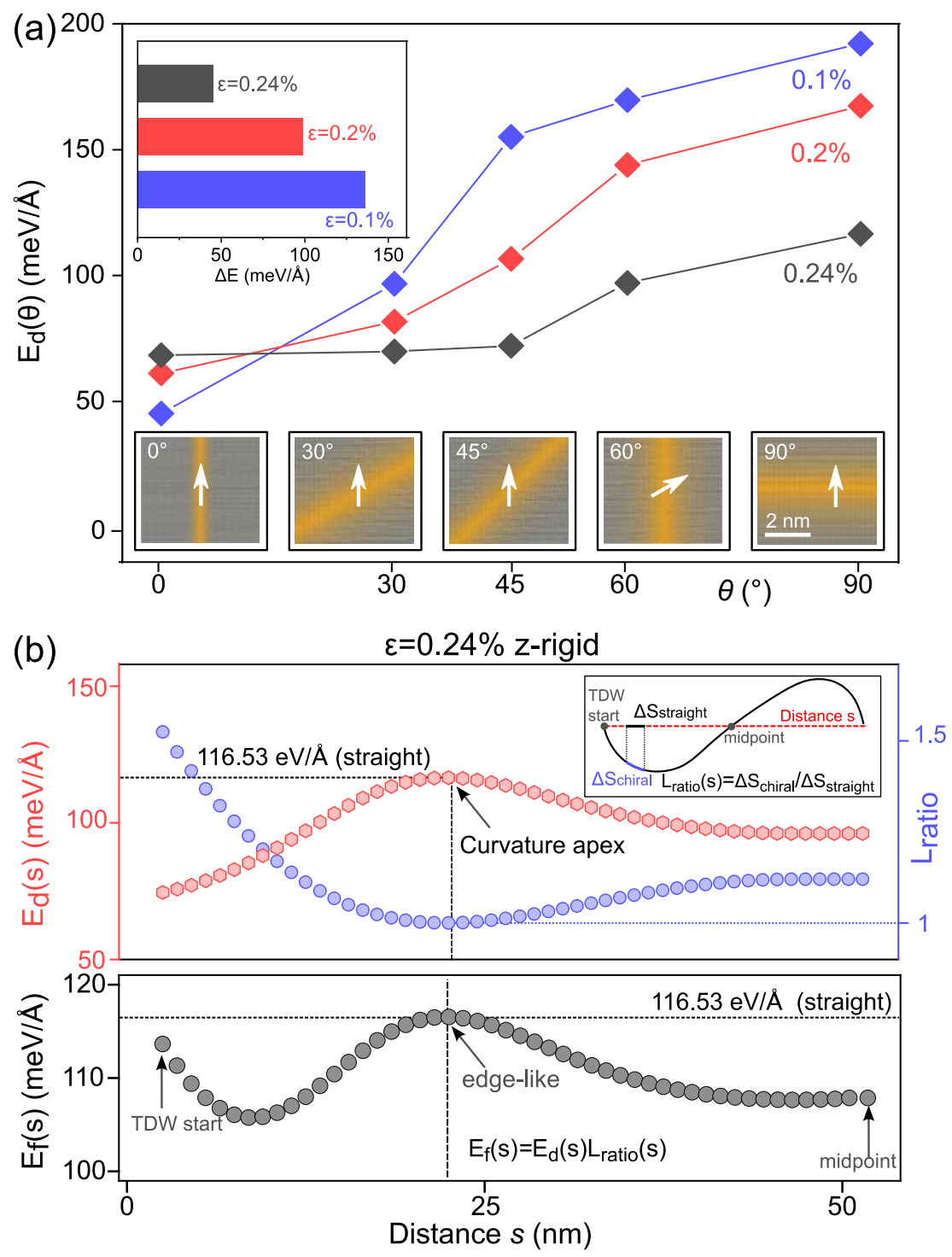

FIG 3. Energetic analysis of curved TDWs. (a) Dislocation formation energy density $E_d(\theta)$ versus TDW orientation angle $\theta$ relative to the Burgers vector for the z-rigid case under different misfit strains. The inset shows the energy difference between screw ($\theta$ = 0°) and edge ($\theta$ = 90°) dislocations. (b) Energetics along a curved TDW at 0.24% misfit strain. Upper: formation energy density $E_d(s)$ and length ratio $L_{ratio}(s)$ (where $\theta$ varies with $s$). Lower: formation energy density $E_f(s) = E_d(s)\times L_{ratio}(s)$.

Directional misfit strain (anisotropic) provides an additional symmetry breaking beyond chirality by making previously equivalent domains unequal in size and shape. Under a biaxial isotropic loading, dual-chiral TDW networks maintain nearly equivalent domain sizes (as seen in Fig. 1(a) and 2(a)). When strain becomes anisotropic, however, the spacing between neighboring TCP rows becomes unequal, and the previously equivalent triangular domains split into two inequivalent groups (Fig. 2(c)). To quantify this effect, we introduce a domain size asymmetry parameter $A_d = 1 - d_1/d_2$, where $d_1$ and $d_2$ denote orthogonal domain dimensions. Under isotropic misfit strain ($\varepsilon_{yy}/\varepsilon_{xx} = 1$ in Fig. 2(c)), $A_d$ remains small (between zero and 0.15) and sporadic across the parameter space (Fig. S5). Under anisotropic misfit strain ($\varepsilon_{yy}/\varepsilon_{xx} \neq 1$), by contrast, domain-size asymmetry becomes pronounced. As summarised in Fig. 3(c), substantial domain-size asymmetry appears only when the directional strain, dual chirality, and sufficient bottom-layer flexibility are all present. A dashed boundary delineates the onset of domain-size asymmetry, separating the symmetric region ($A_d = 0$) from the symmetry-broken region ($A_d \neq 0$). Within the asymmetry regime, most $A_d$ values range from 0.25 to 0.4, indicating a substantial deviation from the symmetric limit.

To understand why TDWs become chiral, we analyze individual TDWs using dislocation theory. Each TDW can be described as a partial dislocation line separating laterally shifted AB and BA domains, so its formation energy density depends on the angle $\theta$ between the TDW line and the Burgers vector [26, 27]. Figure 3(a) summarises this orientation dependence for a rigid bottom layer at three different misfit strains $\varepsilon$ = 0.1%, 0.2%, and 0.24%. In all cases, TDWs with more screw-like character ($\theta \sim 0^{o}$) are energetically favored over edge-like walls ($\theta \sim 90^{o}$). For example, at $\varepsilon$ = 0.1%, the formation energy density decreases from about 200 meV/Å to 50 meV/Å as the wall rotates from an edge-like to a screw-like orientation, which is consistent with elasticity theory [31]. The energy density difference between edge- and screw-like TDWs depends strongly on misfit strain. The difference decreases monotonically from ~147 meV at $\varepsilon$ = 0.1% to ~48 meV at $\varepsilon$ = 0.24% misfit strain. This systematic weakening of the edge–screw energy contrast is consistent with the general trend toward straighter TDWs in Fig. 2 at higher mismatch strain.

To show how this could explain the formation of a chiral network, we examine a representative mono-chiral morphology at 0.24% biaxial misfit strain under the z-rigid constraint [36]. Figure 3(b) provides a quantitative energetic analysis along the TDW (distance coordinate $s$ between the two TCPs). In the upper panel, the red circles (left axis) show the formation energy density $E_d(s)$, obtained by assigning $\theta(s)$ to $E_d(\theta)$. The blue circles (right axis) show the length ratio $L_{ratio}(s)$, as defined in the inset, where the dashed line indicates a straight-TDW reference. The $E_d(s)$ increases as it approaches the curvature apex (increasingly edge-like), whereas $L_{ratio}(s)$ decreases to 1. The lower panel compares the TDW formation energy density $E_f(s) = E_d(s)\times L_{ratio}(s)$. The chiral-TDW formation energy density (grey circles) lies below the straight-TDW reference (black dashed line). The smaller $E_d(s)$, arising from its screw-like character, is sufficient to offset the length increase $L_{ratio}(s)$ with reference to the straight TDW.

To determine how network chirality reshapes low-energy electronic structure, we perform tight-binding calculations under a representative biaxial misfit strain of 0.20% with a z-rigid constraint [36]. Figure 4 compares the local density of states (LDOS) of straight, mono-chiral, and dual-chiral networks under the same misfit strain. Figures 4(a)–4(c) show the LDOS along trajectories crossing TDWs, AB/BA domains, and a TCP, while Fig. 4(d) compares spectra at their centers. Figure 4(e) further resolves the LDOS along opposite edges of representative straight and curved TDWs.

*Contact author: zhe.liu@unimelb.edu.au

For straight TDW networks (Fig. 4(a)), electronic states near the Fermi level are strongly localized at the AA-stacked TCP region, hosting discrete bound states; the characteristic −110 meV doublet is consistent with previously reported values in strained bilayer graphene [19] and resembles the double-peak structures in twisted bilayer graphene, which are associated with Van Hove singularities [32]. The surrounding AB and BA domains remain gapped and display smooth interference patterns, consistent with confined states observed in heterostrained moiré structures [19, 33]. At the TDW itself, a small local gap persists at the wall center, while quasi-one-dimensional boundary modes emerge symmetrically along the two flanking edges, forming nearly mirror-balanced edge channels; notably, the blue (TDW) and light-yellow (AB/BA domain) shaded regions in Fig. 4(a) show that these edge states extend beyond the geometrical TDW edge region into the adjacent AB/BA domains, indicating a broad edge-state distribution.

This electronic landscape is qualitatively reorganized once the network becomes chiral (Fig. 4(b)-(c)). First, the smooth LDOS texture in the AB and BA domains is disrupted by additional curved TDW crossings, resulting in fragmented, step-like modulations rather than the extended interference pattern observed in the straight TDW case. Second, low-energy spectral weight redistributes away from the TCP and toward the walls: the TCP-localized weight weakens and broadens, no longer as tightly confined as in the straight network, while spectral weight correspondingly increases across the TDWs. This redistribution is consistent with the reduction of AA-region size in chiral networks (Fig. 2(b)) and suggests stronger hybridization between the TCP-localized states and the surrounding AB/BA regions.

These trends are corroborated by the local spectra in Fig. 4(d): the pronounced double-peak feature near −110 meV at TCP in the straight network is strongly suppressed and much less distinct in both chiral networks. A comparable suppression of AA-centred peaks has been discussed in twisted bilayer graphene, where the double-peak structure near the Dirac point diminishes or becomes unresolvable at larger twist angles [32, 34, 35]; proposed interpretations include the energy shift and broadening of Van Hove singularities beyond the accessible energy window [34], or a reduction in effective interlayer coupling that leaves spectra more reminiscent of decoupled monolayers [35]. In parallel, spectral weight emerges near the Fermi level along the TDWs, signaling a partial closing of the local gap at wall centers — consistent with the redistribution described above.

Beyond this reorganization of the TCP and domain electronic structure, the clearest manifestation of chirality appears in the TDW edge states (Fig. 4(e)). In straight walls, the two edges remain nearly symmetric, as noted above (Fig. S6); chiral walls, by contrast, exhibit pronounced edge asymmetry and higher density, both of which depend on local curvature. Near TCP regions, low-energy states accumulate preferentially on the outer edge of the curved wall; toward the curvature apex, this asymmetry weakens, while farther along the TDW the preferred edge can even reverse as the local geometry evolves. Curvature therefore acts not as a minor perturbation, but as a primary organizing factor of boundary localization. At the network scale, this asymmetry is further constrained by chiral handedness: in mono-chiral networks, TDWs connect TCPs of identical handedness, forcing inversion of the high-curvature edge between equivalent endpoints, whereas in dual-chiral networks, opposite-handed TCPs impose a local mirror relation that preserves boundary localization on the same physical side near both crossings (Fig. S7). Boundary modes are thus governed simultaneously by local curvature and global network chirality.

In summary, TDW networks are often viewed as collections of equivalent one-dimensional defects whose properties follow from those of an isolated wall, but this picture becomes incomplete once TDWs form a connected, deformable moiré network. Because TDW energy depends on wall orientation relative to the Burgers vector [26,27], curvature can lower the line energy density, yet walls connected through TCPs cannot optimize independently: energy minimization instead occurs collectively under the geometric constraints of the moiré superlattice, allowing relaxation to spontaneously break the symmetry of the straight network and stabilize mono- and dual-chiral geometries as an emergent, collective property of the network rather than of any individual wall. More broadly, related reconstructed networks in moiré and other deformable interfacial systems [3,20–23] suggest that collective symmetry breaking may arise generally from competition between local defect energetics and global geometric constraints.

This geometric symmetry breaking reorganizes the low-energy electronic states, extending the conventional topological framework of moiré electronics: whereas topology guarantees the existence of boundary states [9-11], curvature and chirality determine where and how they localize, establishing network geometry as a programmable, real-space control parameter for electronic behavior, an additional degree of freedom beyond local stacking and valley topology alone.

*Contact author: zhe.liu@unimelb.edu.au

This separation between topological protection and geometric distribution suggests new strategies for engineering anisotropic transport, programmable one-dimensional channels, and curvature-defined electronic architectures in moiré materials, pointing toward a mesoscale design paradigm in which this geometric degree of freedom becomes a fundamental parameter for controlling emergent electronic phenomena.

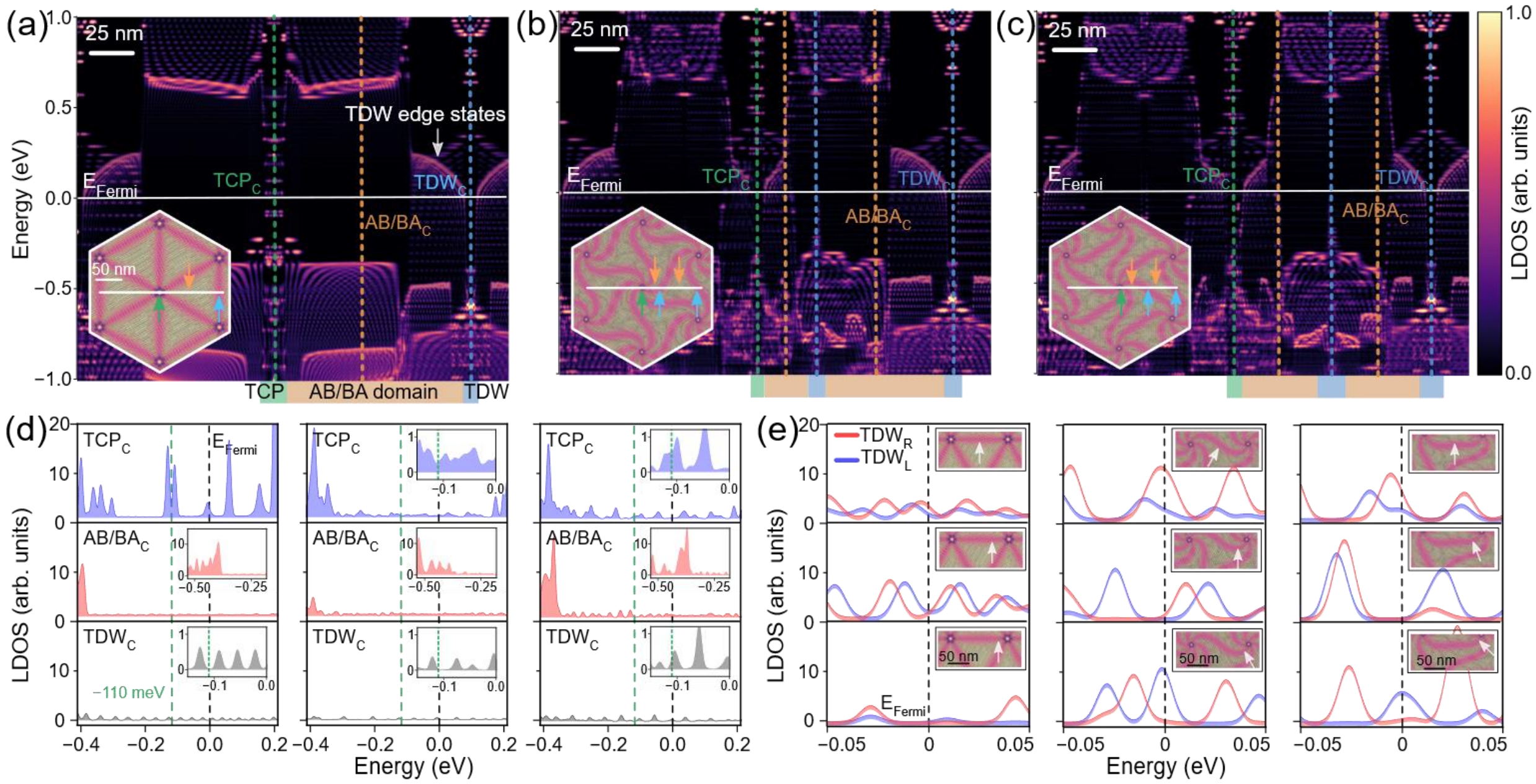

FIG 4. Electronic structures of moiré TDW networks. (a-c). Spatially resolved LDOS along the trajectories (white lines) shown in the insets for straight (a), mono-chiral (b), and dual-chiral (c) TDWs under 0.20% biaxial misfit strain with a z-rigid constraint. (d). LDOS spectra sampled at the corresponding marker positions in (a)–(c). (e). LDOS evolution along the TDW edge away from a TCP outward. The white arrow indicates the sampling direction from the left to the right TDW edge ($TDW_L$ to $TDW_R$).

*Acknowledgments*— This work was supported by the University of Melbourne Early Career Researcher Grant (X.Y.), the Australian Research Council Discovery Project (DP210103888) and Future Fellowship (FT220100149) (J.Z.L.). This work was undertaken with assistance from the National Computational Infrastructure (NCI Australia), an NCRIS-enabled capability supported by the Australian Government. This work was supported by resources provided by the Pawsey Supercomputing Research Centre's Setonix Supercomputer (https://doi.org/10.48569/18sb-8s43), with funding from the Australian Government and the Government of Western Australia.

*Contact author: zhe.liu@unimelb.edu.au

*Contact author: zhe.liu@unimelb.edu.au